\pdfoutput=1
\documentclass[aps,twocolumn,superscriptaddress,notitlepage]{revtex4}
\usepackage{amsmath,amssymb}
\usepackage{xcolor}
\usepackage{graphicx}
\usepackage[colorlinks,linkcolor=blue,urlcolor=blue,citecolor=blue]{hyperref}
\usepackage[normalem]{ulem}
\usepackage{xspace}

\usepackage{pdfpages}

\newcommand{\eref}[1]{Eq.~(\ref{#1})}%
\newcommand{\fref}[1]{Fig.~\ref{#1}} %
\newcommand{\Fref}[1]{Figure~\ref{#1}}%

\begin{document}

\title{Statistics of overtakes by a tagged agent}

\author{Santanu Das}
\affiliation{Raman Research Institute, Bangalore 560080, India}
\author{Deepak Dhar}
\affiliation{Indian Institute of Science Education and Research
(IISER), Pune 411008, India}
\author{Sanjib Sabhapandit}
\affiliation{Raman Research Institute, Bangalore 560080, India}
\date{\today}

\begin{abstract}
We consider a one-dimensional infinite lattice where at each site there sits an agent carrying a velocity,  which is drawn initially for each agent  independently from a common distribution. This system evolves as a  Markov process where a pair of agents at adjacent sites exchange their positions with a specified rate, while retaining their respective velocities, only if the velocity of the agent on the left site is higher. We study the statistics of the 
net displacement of a tagged agent $m(t)$ on the lattice,  in a given duration $t$, for two different
kinds of rates: one in which a pair of agents at sites $i$ and $i+1$ exchange their sites with rate $1$, independent of the velocity difference between the neighbors, and another  in which a pair exchange their sites with a rate equal to their relative speed. In both cases,   we find $m(t)\sim t$  for large $t$. In the first case, for a randomly picked agent, $m/t$,  in the limit $t\to \infty$,  is distributed uniformly on $[-1,1]$ for all continuous distributions of velocities.  In the second case, the distribution  is given by the distribution of the velocities itself, with a Galilean shift by the mean velocity. We also find the large time approach to the limiting forms and compare the results with numerical simulations. In contrast, if the exchange of velocities occurs at unit rate, independent of their values, and irrespective of which is faster, $m(t)/t$ for large $t$  is has a gaussian distribution, whose  width  varies as $t^{-1/2}$. 
\end{abstract}

\maketitle

The phenomenon of overtaking is ubiquitous in nature. It occurs
naturally in all sorts of traffics, ranging from the vehicular traffic on highways~\cite{Chowdhury2000}
to the transport at the molecular scale by motor
proteins~\cite{Schliwa:2003jn, Lipowsky:2006jz}. Animals in groups,
overtake each other to move to a less risky position at the center of
the group~\cite{Ward2016}.  Overtaking also takes place in sedimentation of mixtures with 
polydisperse (different sizes, densities) particles falling (or
rising) through a fluid under gravity ~\cite{Davis85}. In biological evolution, the population sizes
of different genotypes overtake each other depending on their
fitness~\cite{Visser14, Krug03, Jain05, Sire06}. In a completely
different context, the real-time correlation functions in
quantum interacting many body systems may be understood in terms of overtaking dynamics of particles~\cite{Sachdev97,Damle05}.

In spite of its widespread appearances, surprisingly, the statistics of overtakes has not been studied much.  In
this Letter, we  investigate the statistics of
overtakes for a tagged agent in a simple  model of stochastic evolution  of self-driven
agents (e.g., vehicles, molecular motors, etc.) in one dimension.  
In an overtake event,  an agent with a higher velocity crosses another agent with a lower velocity. 
We define
the \emph{net overtakings} for a tagged agent as the total number
of agents that it overtakes   minus the total number of agents that overtake it, in a given duration. We study the probability distribution of fluctuations in this quantity.  Certainly, such statistics can provide useful information about the underlying
traffic.  In particular, in traffic engineering, to obtain flow data,
one uses the
\emph{moving observer method}, where an observer in a test vehicle
moves a fixed distance with a constant speed and counts the number of
vehicles that it overtakes and the number of vehicles that overtake
it~\cite{Chakraborty04}.  In these studies, the fluctuations are usually large, but their  systematic study  is lacking \cite{ARRB}.

\begin{figure}[t!]
\includegraphics[width=.8\hsize]{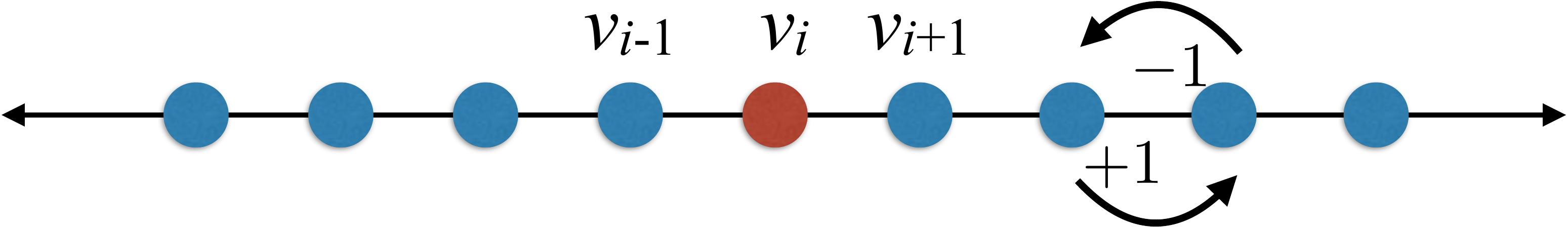}
\caption{\label{model-fig} Each site of a one-dimensional lattice is
occupied by an agent (particle) with a certain velocity, and two
neighboring particles interchange their positions with a specific
rate. During the exchange, the net overtakings of the faster(slower) moving particle increases (decreases) by unity. The
particle in red color is being tagged.}
\end{figure}

In this Letter, we discuss a minimalist model, consisting of a
collection of self-driven agents on a one-dimensional infinite
line. We ignore the actual position of agents, and focus on the relative order. 
Then, to each site $i$ of a lattice, with $i$ ranging from  $-\infty$ to $ \infty$, we associate a real random variable $v_i$, which is the velocity of the $i$-th agent along  the track, starting from some fixed origin.  Each agent is 
assigned a  velocity, at the beginning,  independent of others, from  a   common probability density function (PDF) $\rho(v)$. As agents overtake each other, their relative position with respect to other agents changes, but they retain their respective velocities with them, which are quenched random variables. Therefore, velocity $v_i$ at a given site $i$,  keeps changing as a function of time.  In an exchange
(overtake), the faster particle on the left overtakes a slower
particle (see~\fref{model-fig}) on its right. Therefore the net overtakings in any time interval for a tagged agent equals its shift in position $m$ in that interval on the lattice.

Here we consider two different cases for the exchange rates. 
For an overtake event to take place, the velocity of the
overtaking agent must be greater than that of the overtaken agent. In the first case, we set the rate of exchange for
agents between two neighboring lattice sites $i$ and $i+1$ as
$r=\theta(v_i -v_{i+1})$, where $\theta(v)=1$ for $v>0$ and $0$ for
$v\le 0$, is the Heaviside theta function. In this case, $m(t)$, for a given tagged agent with velocity $v$ increases linearly with time, say as $ m(t) \approx  c(v) t$, and the mean rate of increase of $m$ depends on $v$. This case is related to the totally asymmetric simple exclusion process(TASEP), with infinitely many classes of particles, and has been studied much in literature \cite{Ferrari05, Mountford05, Ferrari91,Ferrari92, Amir11,Angel09}.  Interestingly, if the agent being tracked is picked at random in the beginning, then, the velocity $v$ itself is a random variable, and on averaging over this, we show  below that the  probability distribution of the random variable $c(v)$ is independent of the initial distribution of velocities, and  is uniformly distributed on $[-1,1]$, for all continuous
distributions of the velocities. In the second case, motivated by the consideration of cars on a highway, where the rate of overtaking between two agents is approximately
 proportional to their relative velocities, we take
$r=\theta(v_i -v_{i+1}) \, (v_i -v_{i+1})$.  In this case also,
$m(t)$ again increases linearly with $t$. However, the limiting PDF of
$c=m/t$ is given by the PDF of $v$ itself, with the Galilean shift
$\langle v\rangle$. 
 
We contrast this behavior with what happens to $m(t)$ if the exchange rates $r$ to be independent of the velocities
and uniform everywhere, and adjacent velocities are exchanged at equal rate (chosen as $1$ here), without consideration of which is faster. In this case, it is easy to see that, a tagged agent performs a symmetric random walk (RW), 
 which is well-studied in literature \cite{pearson}.  In this case, the net overtakings $m(t)$, in a given time $t$, has the
diffusive scaling $m(t) \sim \sqrt{t}$ and PDF of the scaled variable $y=m/\sqrt{t}$, in the limit
$t\to\infty$, is Gaussian. 

In the overtaking dynamics, where the exchange occurs only when the faster agent goes to the right, the behavior is very different, as we proceed to show. We can consider both the models described above, by writing in general $r=\theta(v_i -v_{i+1})~ (v_i-v_{i+1})^\alpha$, where $\alpha=0$ and $1$ correspond to the first
and second choices of the rates respectively.  

First consider the case $\alpha =0$.  Our model is equivalent to a  totally asymmetric simple exclusion process (TASEP), with infinitely many classes of particles. 
However, if we consider the motion of a single tagged agent, say starting at the origin, and having the quenched velocity $v_0$, the dynamics of this particle only depends on where the adjacent site has a velocity greater, or less, than $v_0$. Let the density of agents having velocity greater than $v_0$ be $\rho_+(v_0)=\int_{v_0}^\infty \rho(v)$. Then, the motion of this tagged agent is same as that of a single \emph{second class} particle in a TASEP, starting on an initial uncorrelated background of density of first-class particle =$\rho_+(v_0)$, and holes
with density $\rho_-(v_0)=1-\rho_+(v_0)$.  
In this case, the tagged  (second class) particle moves with a velocity
$\bar{c}(v_0)=1-2\rho_+(v_0)$~\cite{Ferrari92}. Evidently,
$\bar{c}(v_0)$ is bounded by $\pm 1$, with $\bar{c}\to \pm 1$ for
$v_0\to\pm\infty$ (or the upper and the lower supports respectively)
and $\bar{c}(v_0^*)=0$ for $\rho_+(v_0^*)=\rho_-(v_0^*)=1/2$. Ignoring
fluctuations around $\bar{c}(v_0)$, the variable $c$ is random through
$c=\bar{c}(v_0)$. Therefore, using    $dc/dv_0= -2\rho'_+(v_0)=2\rho(v_0)$,  we get
\begin{equation}
p(c) = \frac{\rho(v_0(c))}{|dc/dv_0|}=\frac{1}{2}.
\label{limit-dist}
\end{equation}
Thus $c=m(t)/t$, in the limit $t\to \infty$, is uniformly distributed
on $[-1,1]$, for all continuous distributions $\rho(v)$, as claimed above.  

A somewhat similar result was obtained earlier for TASEP with the step initial condition, and it was shown that a second class particle starting at the step,  acquires a limiting speed that 
 is uniformly distributed on $[-1,1]$ ~\cite{Amir11, Angel09}. The corresponding result for our case is that when the initial velocities  in our model with $\alpha=0$ are  in the descending order,  $v_i > v_{i+1}$ for all $i$,  an agent picked at random has a limiting speed,  uniformly
distributed on $[-1,1]$.  

\begin{figure}
\includegraphics[width=.9\hsize]{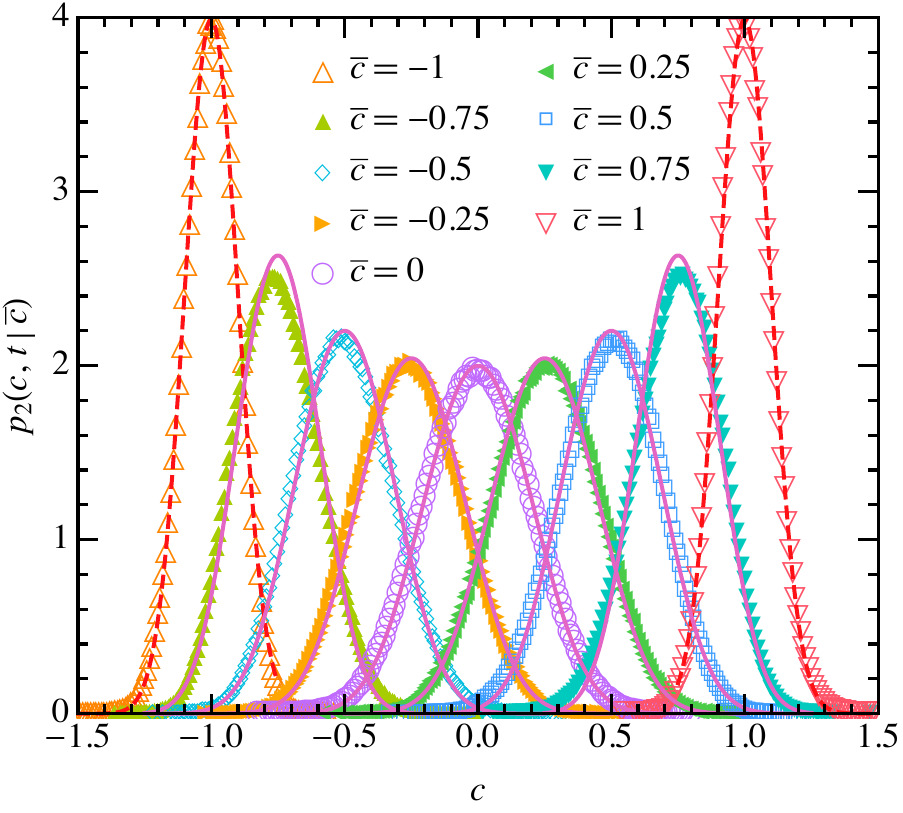}
\caption{\label{fig2} (Color online). The points are numerical simulation results for the the conditional PDF of scaled net overtakings $c = m(t)/t$ for given   $\bar{c}$, 
for our first choice of the overtaking rate $r = \theta(v_i - v_{i+1})$. The initial velocities are drawn from uniform distribution on $[-1,1]$. The dashed lines plot two Gaussian distributions centered around $\bar{c}=\pm 1$ respectively, with a variance $t^{-1}$. The solid line plots \eref{PraSpohn} for $\bar{c}=0, \pm0.25, \pm 0.5 ~\text{and}~ \pm 0.75$. In all the cases, $t=100$.
}
\end{figure}

One can also study fluctuations around the mean drift velocity $\bar{c}(v_0)$, for a given $v_0$, at large, but finite, times $t$.     The PDF of $c$ at any  time $t$ (for any $\alpha$) is given by
\begin{equation}
p(c,t) = \int_{-\infty}^\infty p_1(c, t|v_0)\, \rho(v_0)\, dv_0,
\label{approach1}
\end{equation}
where $p_1(c,t|v_0)$ is the conditional PDF for a given
$v_0$. For $\alpha=0$, as  $p_1(c,t|v_0)$ is expected to depend on
$v_0$ only through $\rho_+(v_0)$, or equivalently
$\bar{c}(v_0)$,  making a change of  variable from $v_0$ to
$\bar{c}$,   
 eliminates  $\rho(v_0)$  completely   from \eref{approach1},
\begin{equation}
p(c,t)=\frac{1}{2}\int_{-1}^1 p_2 (c,t|\bar{c})\, d\bar{c},
\label{approach2}
\end{equation}
where  $p_2(c,t|\bar{c})$ is the conditional PDF for a given $\bar{c}$.
Thus, not only the limiting distribution, but also the $p(c,t)$ at all
time is independent of the velocity distribution $\rho(v)$. Note that,
while obtaining \eref{limit-dist}, we have taken
$p_2(c,t|\bar{c})=\delta(c-\bar{c})$ by ignoring the fluctuations.

If we ignore the correlations between the jumps of the tagged particle at different times, then the typical fluctuations of $c$ around the mean velocity $\bar{c}(v_0) $, at the scale of the standard deviation $\sigma_t=\sqrt{\langle [c-\bar{c}(v_0)]^2\rangle} = t^{-1/2}$,   are Gaussian.
While this simple RW description holds good at an early time,  the correlations between jumps build up at later times~\cite{SM}.  Eventually,  it crosses over to $\sigma_t \propto \chi^{1/3} t^{-1/3}$
behavior~\cite{vanBeijeren85, Prahofer02, Ferrari06} 
with  $\chi=\rho_+(v_0)
[1-\rho_+(v_0)]=(1-\bar{c}^2)/4$, and typical fluctuations are
described by~\cite{Prahofer02, Ferrari06}
\begin{equation}
p_2(c,t|\bar{c}) \simeq \frac{1}{4} (2\chi^{1/3} t^{-1/3})^{-1}\,
G_\mathrm{scaling}\bigl([c-\bar{c}]/[2\chi^{1/3} t^{-1/3}]\bigr),
\label{PraSpohn}
\end{equation}
where $G_\mathrm{scaling}(w)$ is the scaling function associated with the spatio-temporal two-point correlation function of the TASEP with the Bernoulli product measure initial condition~\cite{Prahofer02, Ferrari06, BaikRains, Prahofer04, Prahofer03}.  The
crossover time 
$t_* \propto \chi^{-2}$.

\begin{figure}
\includegraphics[width=.95\hsize]{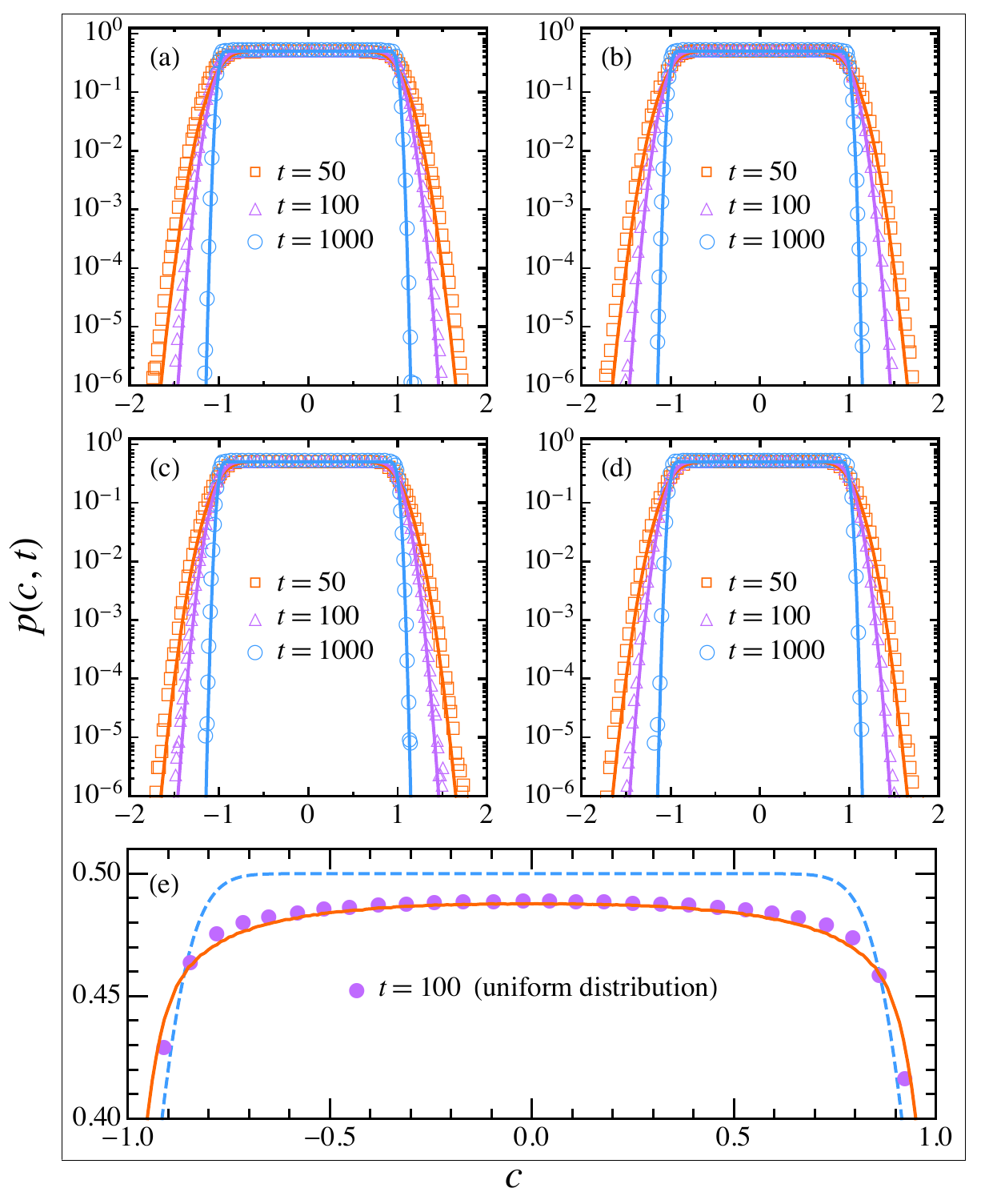}
\caption{\label{fig-case2} (Color online). In (a), (b), (c), and (d): The points are numerical
simulation results for the PDF of the scaled net overtakings
$c=m(t)/t$ at different times, for our first  choice of the
overtaking rate $r=\theta(v_i-v_{i+1})$, where the initial velocities are chosen from (a) uniform, (b) Gaussian, (c) exponential, and (d) power-law distributions.  
 The solid lines plot \eref{asymptotic-case2}. In (e): The points  are from a numerical simulation where the initial velocities are chosen from a uniform distribution on $[-1,1]$ and $t=100$.  The solid line plots   \eref{approach2} computed by using \eref{PraSpohn} and the dashed line plots \eref{asymptotic-case2} for $t=100$.
}
\end{figure}

At large times, since, $p_2(c,t|\bar{c})$ in \eref{approach2} is peaked sharply around
$\bar{c}$, the correction to \eref{limit-dist} near the
edges $c=\pm 1$, comes from $\bar{c}\to \pm 1$ respectively.  In this
case, $t_*$ diverges.  Hence, we can use the Gaussian form [see \fref{fig2}] 
$p_2(c,t|\bar{c})\simeq \exp(-t[c-\bar{c}]^2/2)/\sqrt{2\pi
t^{-1}}$ of  the RW picture in \eref{approach2}. This yields 
\begin{equation}
p(c,t)\simeq \frac{1}{4} \left[
\text{erf}\left(\frac{(c+1) \sqrt{t}}{\sqrt{2}}\right)-
\text{erf}\left(\frac{(c-1) \sqrt{t}}{\sqrt{2}}\right)\right],
\label{asymptotic-case2}
\end{equation}
where $\text{erf}(x)=(2/\sqrt{\pi}) \int_0^x e^{-y^2}\, dy$ is the error function.  
We compare this form with numerical results in \fref{fig-case2} and find very
good agreement. The agreement becomes even better, if  the above Gaussian approximation is replaced by 
the large deviation form  of the distribution for the RW~\cite{SM}. 
The finite large time correction for the central region around $c=0$ can be computed by
using \eref{PraSpohn} [see \fref{fig2}] in \eref{approach2} [see~ \fref{fig-case2}(e)].

Let us next consider the case
$\alpha=1$.  For this, our model, in fact, corresponds to the
infinite-species Karimipour model~\cite{Karimipour99}.  As in the $\alpha=0$ case, here also,  from numerical simulation we find that~\cite{SM},  
 the motion of the tagged agent with a given velocity $v_0$ can be well-approximated  by a
RW for  $t \ll t_\#(v_0) $,  where the characteristic time $t_\#(v_0)$ increases with $v_0$ going further away from the mean value  $\langle
v \rangle =\int_{-\infty}^\infty v\rho(v) dv$. 
The RW  jumps to the right with the rate
$\rho_R(v_0)=\int_{-\infty}^{v_0} (v_0-v)\, \rho(v)\, dv$ and to the
left with the rate $\rho_L(v_0)=\int_{v_0}^{\infty} (v-v_0)\, \rho(v)\,
dv$.  Therefore,  the drift velocity $\bar{c}(v_0)
= \rho_R(v_0) -\rho_L(v_0) = v_0 - \langle v \rangle$  and the standard deviation $\sigma_t=\sqrt{[\rho_L(v_0)+ \rho_R(v_0)]/t}$.
Even for $t \gg t_\#(v_0)$, from our simulation, we find that the typical fluctuations around its mean  $\bar{c}(v_0) t$ are well-described by a Gaussian distribution, albeit with a  standard deviation $\sigma_t \propto t^{-1/2} s(t)$. The factor $s(t)$ gives the correction to the standard RW result, which may be either a very small power-law $s(t)\sim t^{0.06\dots}$ or a logarithmic correction $s(t) \sim (\ln t)^\gamma$, and 
it is difficult to distinguish between the two behaviors based on our numerics~\cite{SM}  --- as is often the case with marginal corrections~\cite{Hoef91, Lowe95, Isobe08, Krug18}.
Such logarithmic corrections have been found earlier 
in  two-dimensional driven diffusive systems~\cite{vanBeijeren85, Landim04, Yau04, Quastel13} as well as for $1+1$ dimensional interface with cubic nonlinearity~\cite{Derrida91, Devillard92, Paczuski92, Binder94}.

\begin{figure}
\includegraphics[width=.95\hsize]{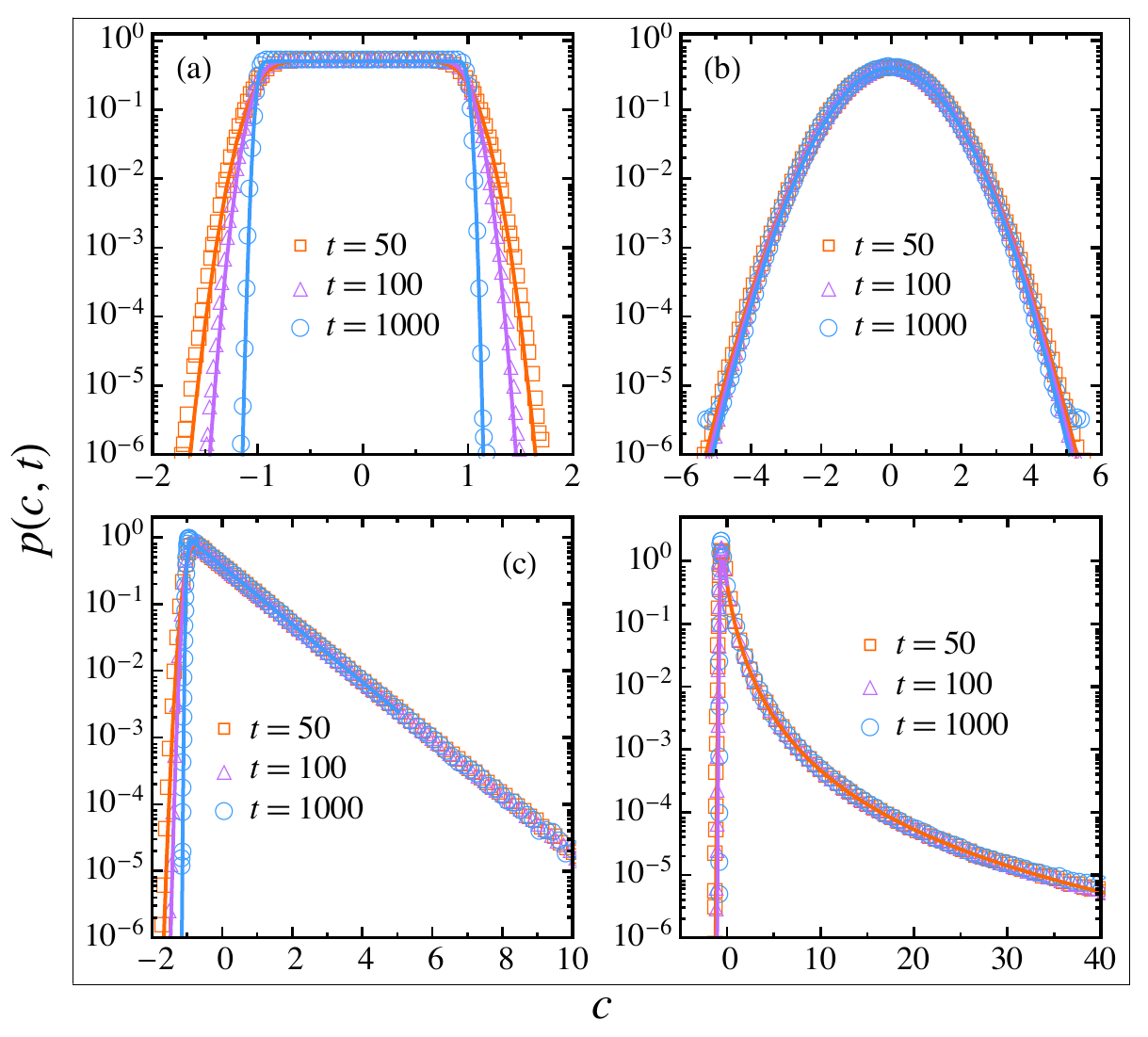}
\caption{\label{fig-case3} (Color online). The points are numerical
simulation results for the PDF of the scaled net overtakings
$c=m(t)/t$ at different times, for the second choice of the
overtaking rate $r=\theta(v_i-v_{i+1}) (v_i-v_{i+1})$, where the initial velocities are chosen from (a) uniform, (b) Gaussian, (c) exponential, and (d) power-law distributions.  
The solid lines plot the
theoretical results obtained from \eref{case3-asymptotic}.
}
\end{figure}

In the limit $t\to\infty$, ignoring the fluctuations around
$\bar{c}$ gives the limiting PDF $p(c) = \rho(c+\langle v\rangle)$, as
announced above. The shift of the PDF by the mean is easily
understood, as the overtaking dynamics depends only on the velocity
differences. For the approach to this limiting distribution, we note that as in the $\alpha=0$  case, the contributions to the large $|c|$ tails of $p(c,t)$ comes from  large $|v_0 -\langle v\rangle |$ behavior of $p_1(c,t|v_0)$ in \eref{approach1}. Since $t_\#$ is large for large $|v_0 -\langle v\rangle |$,  the tails of $p(c,t)$ 
can be computed by using a Gaussian distribution  with mean $v_0-\langle v\rangle$ and variance  $[\rho_L(v_0)+ \rho_R(v_0)]/t$ for  $p_1(c,t|v_0)$ in \eref{approach1},
\begin{align}
p(c,t)\simeq \int_{-\infty}^\infty
 dv_0\,  \rho(v_0)\,
\frac{\sqrt{t}}{\sqrt{2\pi  [\rho_L(v_0)+ \rho_R(v_0)]}} \notag\\
\times 
\exp \left(-\frac{t[c + \langle v\rangle - v_0]^2}{2[\rho_L(v_0)+ \rho_R(v_0)]}
\right).
\label{case3-asymptotic}
\end{align}
Evidently,  $p(c,t)$ now depends on the form of $\rho(v_0)$ and the
integral has to be carried out separately for each case.   \Fref{fig-case3} shows very good agreement between \eref{case3-asymptotic} and numerical simulation results for four different choices of $\rho(v_0)$.

It is interesting to note that the above RW picture for the overtaking
dynamics, with the hopping rates $\rho_L (v_0)$ and $\rho_R(v_0)$, is
exact for a model, known as the Jepsen gas~\cite{Jepsen65}, where the
particles are non-interacting, distributed uniformly with unit density
on the one-dimensional (continuous) infinite line, and move
ballistically with their velocities drawn independently for each
particle from $\rho(v)$. This, for example, models the dynamics of
cars on multi-lane highways at low densities.

As a side remark, in the case when the random velocity
$v$ takes only two distinct values,   clearly, the value of $\alpha$ does not matter anymore, and the model corresponds to TASEP with identifying agents carrying one type of velocity as particles and the agents carrying the other type of velocity as holes. 
In this case, the tagged particle displacement $m(t)$ depends
on the initial condition~\cite{vanBeijeren91, Majumdar91, Borodin07, Imamura:2007hp}.
In particular, for independent and identically drawn initial velocities, 
 a tagged particle performs a totally asymmetric
RW in continuous time, where the fluctuations
about the mean displacement  are Gaussian and  grows diffusively in time ~\cite{deMasi85, Kunter85, Kipnis86, Kesten}.

In conclusion, we have found two categories of overtaking
behavior.  In these cases, the net number of overtakes  grows as $t$. We have also obtained the limiting distribution of the
time-averaged overtaking rate, defined by the total number of net
overtakings in a given time period divided by the total time, as well
as the approach to the limiting distributions.

There are several interesting open directions for future
research. First and foremost one is of course to analyze real
data. Another question is whether there are other classes, and if any,
how to identify them.  Third, here we have studied only a single time
property. However, one can study correlations between overtakes at
different times or the overtaking dynamics itself as a process.  Here,
it is somewhat assumed that the density of agents is homogeneous in
the real space, so that velocity is the only relevant variable for
overtaking. One can explore the effect of inhomogeneity by considering
a dilute case, where a finite number of sites, chosen randomly with a
given density, are not occupied by agents (equivalently, occupied
by agents having zero velocity).  The simple picture presented in
this Letter can serve as a stepping stone for future studies.
The model on a finite line is also of interest. In this case, there are important end effects, and there are shock waves that start at the ends and travel inwards, and determine the qualitative behavior in the region deep inside for times of order of the system size. These will be discussed in a future publication.

\widetext

\vskip3cm 
{\bf \large Supplemental Material included from the next page.}

\newpage

\pagestyle{empty}

\hskip-2.5cm\includegraphics[page=1]{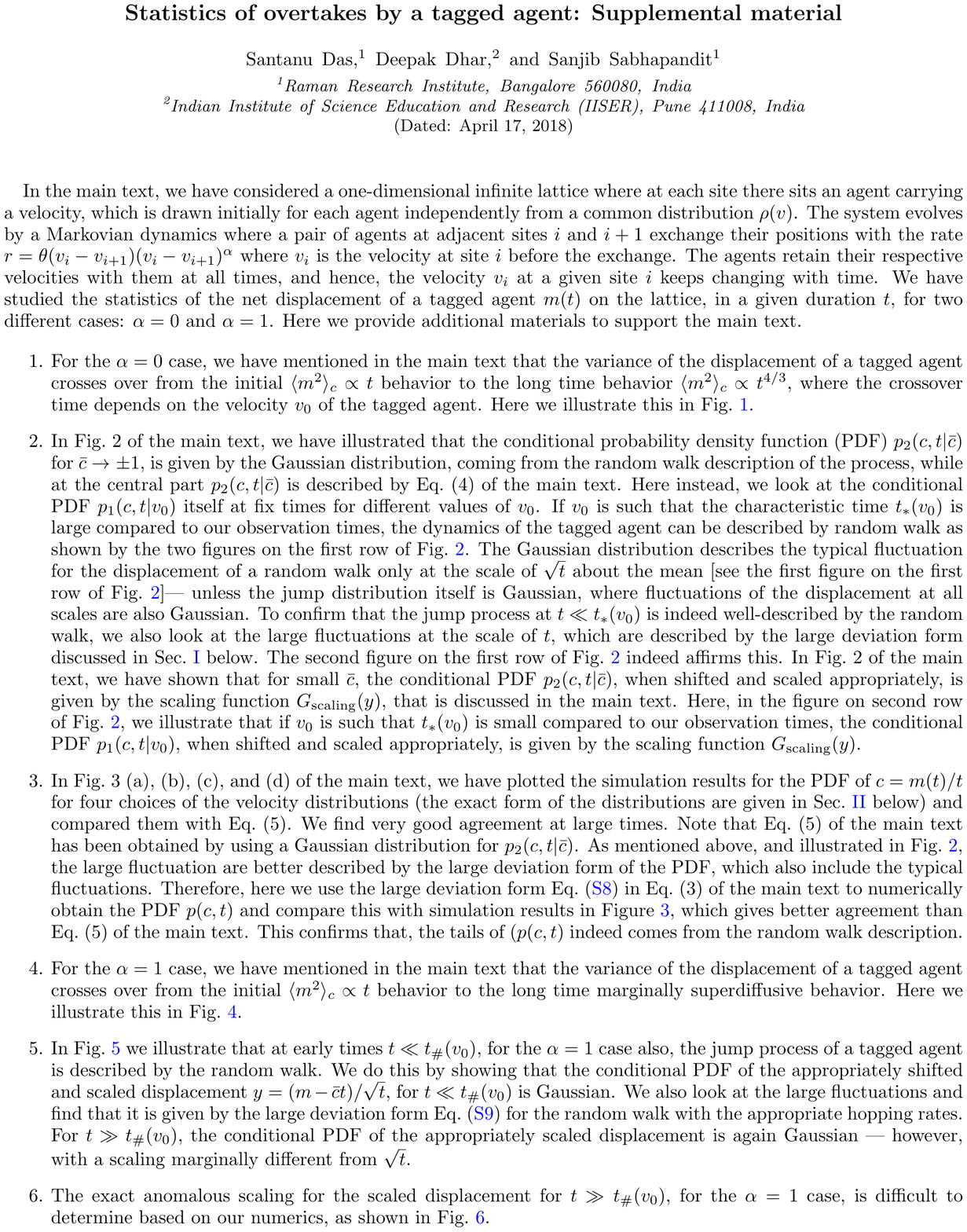}  

\hskip-2.5cm\includegraphics[page=2]{supplmental.pdf}  

\hskip-2.5cm\includegraphics[page=3]{supplmental.pdf}  

\hskip-2.5cm\includegraphics[page=4]{supplmental.pdf}  

\hskip-2.5cm\includegraphics[page=5]{supplmental.pdf}  

\hskip-2.5cm\includegraphics[page=6]{supplmental.pdf}  

\hskip-2.5cm\includegraphics[page=7]{supplmental.pdf}  

\hskip-2.5cm\includegraphics[page=8]{supplmental.pdf}  

\hskip-2.5cm\includegraphics[page=9]{supplmental.pdf}  

\hskip-2.5cm\includegraphics[page=10]{supplmental.pdf}  

\end{document}